\documentclass[preprint,aps,showpacs]{revtex4}

\usepackage{graphicx}
\usepackage{dcolumn}
\usepackage{bm}

\begin{document}

\title{Direct observation of the flux-line vortex glass phase in a\\ type II superconductor}

\author{U. Divakar, A.J.~Drew, S.L.~Lee}
\affiliation{School of Physics and Astronomy, University of St.~Andrews,
St.~Andrews, Fife KY16 9SS, UK.  }
\author{R.~Gilardi, J.~Mesot}
\affiliation{Laboratory for Neutron Scattering, ETH Z\"urich and PSI 
Villigen, CH-5232 Villigen PSI, Switzerland}
\author{F.Y.~Ogrin}%
\affiliation{Department of Physics, University of Exeter, Exeter EX4 4QL, UK}
\author{D. Charalambous, E.M.~Forgan}
 \affiliation{School of Physics and Astronomy, University of Birmingham,
 Birmingham B15 2TT, UK.}
\author{G.I.~Menon}
 \affiliation{The Institute of Mathematical Sciences, C.I.T. Campus, Taramani, Chennai 600 113, India}
 \author{ N.~Momono, M.~Oda}
 \affiliation{Department of Physics, Hokkaido University, Sapporo 060-0810, 
Japan}
 \author{C.D.~Dewhurst}
 \affiliation{Institut Laue-Langevin, 6, rue Jules Horowitz, B.P. 156 - 38042, Grenoble Cedex 9, France}
 \author{C.~Baines}
  \affiliation{Laboratory for Muon Spin Spectroscopy, PSI Villigen, CH-5232 Villigen PSI, Switzerland
}
\date{\today}

\begin{abstract}
The order of the vortex state in
La$_{1.9}$Sr$_{0.1}$CuO$_{4}$ is probed
using muon spin rotation and small-angle neutron
scattering.  A transition from a Bragg glass to a
vortex glass is observed, where the latter
is composed of disordered vortex lines. In the
vicinity of the transition the microscopic
behavior reflects a delicate interplay of
thermally-induced and pinning-induced disorder.
\end{abstract}

\pacs{PACS numbers: 74.25.Qt,74.72.Dn, 76.75.+i, 61.12.Ex}
\maketitle

It has been acknowledged for many years
\cite{Larkin79}  that the long range
translational order of the Abrikosov vortex
crystal in type II superconductors should be
disrupted by  arbitrarily small amounts of
disorder. The current theoretical perspective is
that weak disorder results in a `Bragg glass'
(BG), which retains the topological order of the
flux line lattice (FLL) but yields broadened
diffraction peaks \cite{Giamarchi94}. Strong
disorder should result in the topologically
disordered `vortex glass' (VG)
\cite{Giamarchi94}.  The BG concept accounts for
the observation of an apparently ordered FLL in
many systems (e.g. \cite{Cubitt93,Mesot02}) and
further evidence for this phase in an isotropic
superconductor has recently been obtained using
small angle neutron scattering (SANS)
\cite{Klein01}.  At high magnetic fields and low
temperatures, where the effects of disorder are
stronger, the BG should be unstable to the
formation of the VG.

In contrast to the enormous wealth of
experimental work on vortex matter derived  from
macroscopic measurements
\cite{blatterbible,Zeldov,Schilling,Pardo}, there
have been relatively few  microscopic studies.
Of these techniques SANS and muon spin rotation
($\mu$SR) measurements provide unique information
concerning  the local and long range  order of
the vortex system (e.g.
\cite{Cubitt93,Mesot02,Klein01,review}).  Here we
report  muon-spin rotation measurements on
underdoped La$_{1.9}$Sr$_{0.1}$CuO$_{4}$ (LSCO)
which provide  the first unambiguous experimental
evidence for the transition from an ordered phase
(BG)  to a VG in a system of well-coupled vortex
lines.

The importance of  LSCO as a system in which to
study vortex matter, and in particular the BG to
VG transition, can be understood in terms of two
key parameters, namely the  superconducting
penetration depth $\lambda_{ab}$  and the
anisotropy parameter $\gamma$.  The former
determines the ability of the supercurrents in
the copper-oxide sheets ({\em ab}-planes) to
screen externally applied magnetic fields, while
the latter quantifies the uniaxial anisotropy of
the penetration depth.  In addition,
the Josephson length
$\lambda_{J}=\gamma s$ (where  $s$ is the spacing
of the copper oxide  planes), determines the
effectiveness of currents, tunnelling  between
the conduction planes,  in maintaining the stiffness
of a vortex line  \cite{Blatter_em} .  While
higher values of $\lambda_{ab}$, $\gamma$
increase the susceptibility of the vortices to
disordering, it is the ratio
$\lambda_{J}/\lambda_{ab}$ which determines the
dimensionality of the vortex system. For example,
in  optimally doped
YBa$_{2}$Cu$_{3}$O$_{7-\delta}$  (YBCO) $
\lambda_{J} / \lambda_{ab} \ll 1$, so the
vortices resemble the rigid rods  of flux found
in conventional isotropic  superconductors. In
contrast, for optimally-doped
Bi$_{2}$Sr$_{2}$CaCu$_{2}$O$_{8+\delta}$
(BSCCO)   $\lambda_{J}/\lambda_{ab}
\stackrel{_{>}}{_{\sim}}1$; consequently the
vortex system is quasi-two dimensional. As we
demonstrate here,  the novelty of underdoped
LSCO is that the material parameters of
($\lambda_{ab} \sim 3000$  \AA\ and
$\lambda_{J}  \sim 250$ \AA\ ) give rise to a
system of fairly rigid vortex {\em lines}  which
are nonetheless highly susceptible to transverse
fluctuations.

Our $\mu$SR measurements were performed using the
GPS spectrometer on the $\pi$M3 beamline at the
Paul Scherrer Institut, Switzerland. The field
was applied parallel to the $c$-axis of the
crystal, and the  muon spins were  initially
polarised nearly perpendicular to the applied
field. The muons implant in the sample at random
positions with respect to the vortices, and each
precesses at a rate $\omega = \gamma_\mu  B(r)$,
determined by the local field $B(r)$ and the
gyromagnetic ratio of the muon. The muon spin
precession is monitored via the positrons which
are emitted preferentially along the muon-spin
direction during  the  decay of the muon
\cite{review}. The  precession signal, averaged
over $\sim 10 ^{7}$ muons, reveals the
probability distribution $p(B)$ of the spatial
variation $B(r)$.  The sample was a single
crystal of length 20~mm and diameter 5~mm, with
the $c$-axis perpendicular to the axis of the
cylinder. The superconducting transition
temperature of the sample was  $T_{c} = 29$~K
with a width (10\% - 90\%) $\Delta T = 1.5$~K,
indicating the high quality of the sample.  All
$\mu$SR measurements were made after cooling in
an applied field in order to obtain a uniform
flux density across the sample.  

\begin{figure}[thb]
\includegraphics[width=16cm]{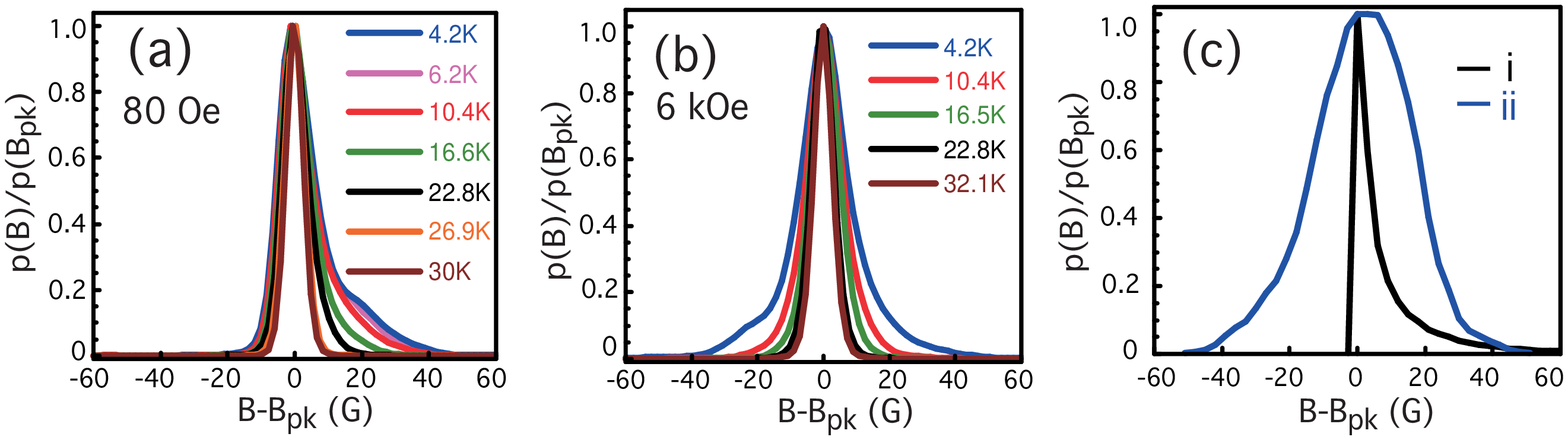}
\caption[]{\label{fig1} (Color)
a) Temperature-dependent probability distributions  of the 
internal field,  $p(B)$, due to the presence of the flux lines  at 
applied fields of a) 80 Oe and b) 6 kOe.  The curves are normalised  at the mode of 
the  distribution to allow easy comparison of the shapes.  c) Field distributions derived from Monte Carlo simulations of the vortex state at 6~kOe  (see text) for the case of:  i) an ideal vortex lattice ; ii) a VG phase with a transverse 
correlation length of about  six lattice spacings. The latter also includes a small contribution due to the effects of thermal narrowing (as observed experimentally). The curves are again normalised  to the mode of  the  distribution.  Note  the increase in width and symmetry of the lineshape due to the increased disorder.}
\end{figure}

In a $\mu$SR
experiment the presence of a vortex line lattice
is revealed by a $p(B)$ which has a distinct
signature, most notably a peak followed by a
tail  extending towards fields higher than the
mean, reflecting the small volume of the core
region of the vortices. The peak at the  mode of
the distribution represents the high probability
density of the field value at the centre of a
line connecting two vortex cores \cite{review}.
In Fig.~\ref{fig1}a are shown probability
distributions  measured in LSCO at low field,
where those taken at the lowest temperatures
correspond most closely to the lineshape
expected from an ideal lattice.  One of the
calculated lineshapes given in  Fig.~\ref{fig1}c
is also derived from an ideal vortex lattice,
where for clarity the rounding effects of
instrumental resolution and nuclear dipolar
broadening, visible in the experimental data of
Fig.~\ref{fig1}a,
 have been  omitted.  Departures from this ideal
form may be used to deduce changes in the
morphology of the vortex state \cite{review}.
The muon probe measures the {\em local} field,
which is determined by contributions from
vortices only within $\sim \lambda_{ab}$  of the
muon; hence, the general form of  $p(B)$ is
remarkably robust to changes in {\em long range}
order  \cite{Brandtlowtemp}.
In contrast, the technique can be highly
sensitive to changes in the {\em local}
environment due to thermally-induced or
pinning-induced distortions \cite{review}, which
allows us to observe the transition from the
ordered state to the VG  phase in LSCO.

The main result of this paper can be appreciated
immediately  by comparing the  normalised
experimental lineshapes of Fig.~\ref{fig1}a and
~\ref{fig1}b, measured at 80~Oe and 6~kOe
respectively.  Referring first to the lowest
temperature lineshape at each field, it is clear
that they are very different. While the 80~Oe
distribution has the characteristics of an ideal
vortex lattice described above,  the 6~kOe signal
is highly symmetric, indicating a strong
departure from an ideal vortex lattice  or Bragg
glass arrangement.  For comparison, in
Fig.~\ref{fig1}c we include the $p(B)$ derived
from Monte Carlo simulations of  a perfect
triangular Abrikosov lattice and also for a
vortex glass structure  having
short-ranged translational correlations of the
order of  about six inter-vortex spacings.   The
simulations are based on one possible mechanism
for the loss of long range order recently
discussed by one of us \cite{Menon02}.  In this
scenario, the system undergoes a transition  from
a BG phase  to a multi-domain glass comprising a
size distribution of domains within which the
vortices are locally ordered.  Increasing the
field leads to a rapid fall in the average domain
size just above  the BG phase.   Note that the
distribution function obtained for the
multi-domain structure is far more symmetric as
well as broader than is the case for the perfect
lattice, as is found experimentally in the data
at 6~kOe (see Fig.~\ref{fig1}b).

\begin{figure}[thb]
\includegraphics[width=16cm]{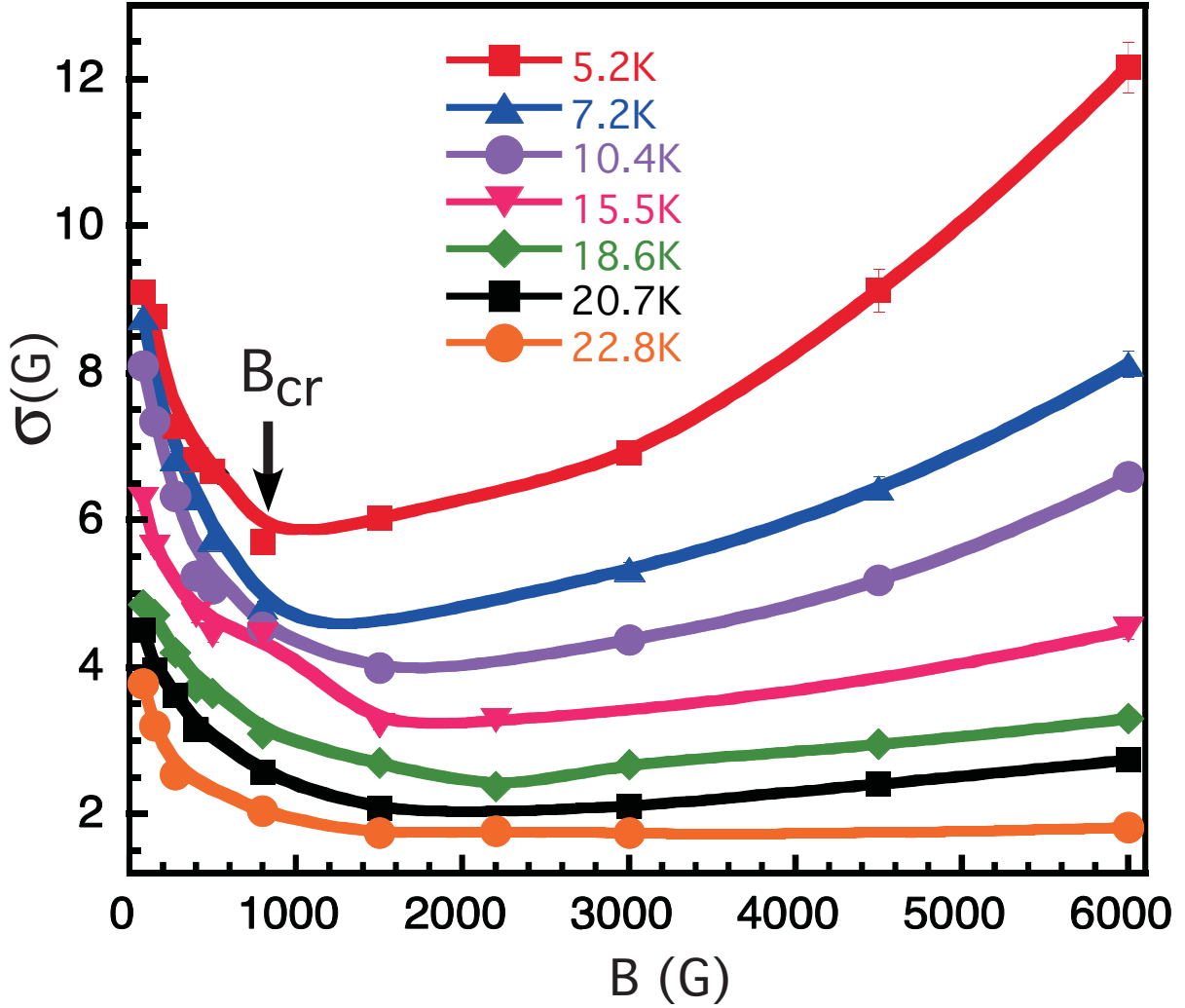}
\caption[]{\label{fig2} (Color)
a) The variation with field of the line width $\sigma$ at 
selected temperatures. The initial reduction with increasing field corresponds to  the field dependent influence on the field distribution of  both 
thermal fluctuations and increasing elastic vortex interactions.  Above $B_{cr} \sim 1$~kOe, the width 
dramatically rises due to the onset of static disorder in the VG phase.  Note that the high-field 
dependence has almost disappeared  by a temperature of 23~K, and disappears completely  at  the irreversibility line (see Fig.~\ref{fig4}).   }
\end{figure}

Further conclusive evidence for the transition
from a BG to a VG phase comes from examining
the width of the lineshapes $\sigma$, given by
the  square root of the second moment of $p(B)$.
This quantity is plotted in Fig.~\ref{fig2} as a
function of applied field, for  a range  of
temperatures. We first draw attention to the
data at the lowest temperatures, where it
can be seen that the signal measured at 6~kOe is
considerably broader than that at 80~Oe. It has
been shown theoretically
\cite{Brandtpancakes,Menon99} that such a {\em
broadening} of the signal from the vortex lattice
can only arise  from {\em static}  disorder in
the positions of  vortex {\em lines}  within a
plane perpendicular to the field \cite{caveat}.
This arises due to positional disorder of the
vortices, which
gives rise to regions with field values  both
higher and lower  than in the well-ordered
lattice
\cite{Brandtlowtemp,Brandtpancakes,Menon99,Brandt88}.
Conversely, Brandt  showed that  in a system
composed of two-dimensional `pancake-vortex'
strings \cite{Brandtpancakes}, short-wavelength transverse
fluctuations along the
field direction  would always lead to a {\em
narrowing} of the field distribution $p(B)$
\cite{Brandtpancakes}, and indeed  this was the
situation found experimentally in the very
anisotropic (quasi-two-dimensional) BSCCO
\cite{lee93}. Thus it is clear that in LSCO we
are dealing with a highly disordered vortex {\em
line} arrangement, which we identify with the VG
phase.

We now turn our attention to the temperature
dependence of the lineshapes of
Fig.~\ref{fig1}a,  where it can be seen 
increasing temperature truncates
the high-field tails.  This reflects the
increasing amplitude of thermally-induced
fluctuations $\langle u^2 (T) \rangle ^{1/2}$  of
the vortex positions, on a timescale faster than
the characteristic muon sampling time
\cite{Song93}. The muons thus experience a
time-averaged field distribution in which  the
high-field values  arising from close to the
vortex cores are smeared out.  Compared to  those
at low temperature,  the lineshapes are narrowed.
Such narrowing
arises both from the increase in $\lambda
(T)$  and  $\langle u^2 (T)\rangle ^{1/2} $. This
may be described by \cite{lee95,Song93}:
\begin{equation}\sigma ^{2} (T) = \langle B
\rangle ^2\sum_{\vec{\tau} \neq 0}
\frac{e^{-\tau^2\langle u^2 \rangle
/2}}{(1+\lambda^2(T)\tau^2)^2},\label{D-W}\end{equation}
where the $\vec{\tau} $ are reciprocal lattice
vectors, multiples of $ \pi/d$, and  $d \sim
\sqrt{\Phi_o/B}$ is the plane spacing of the
vortex lattice. At a given temperature, the width
$\sigma$ decreases as the field is increased,
mainly due to an increase in the ratio $\langle
u^2 \rangle ^{1/2}/d $. This is in contrast to
the case of  an ideal static  vortex lattice,
where $\sigma$ is {\em independent}  of field in
the range  $H_{c_{1}}<H\ll H_{c_{2}}$.  In
Fig.~\ref{fig2} one observes a reduction in
width  with increasing field up to fields of
around 1~kOe.  While  the observed changes of the
lineshapes show that this must at least partly be due to
these dynamical effects, simulations indicate
that the magnitude of this narrowing cannot be
entirely attributed to this mechanism. The
broadening  towards lower fields must also
include a contribution  from  an increase in the
static disorder  due to the reduction of elastic
interactions between vortices in this very dilute
vortex state. Indeed, it is only at fields above
a few hundred Oersted  that the vortex separation
becomes less than the penetration depth in this
system, whereby the shear modulus for the lattice
increases significantly  \cite{Blatter_em}. Thus
below the BG-VG transition one would also expect
the linewidth in this system to increase with
decreasing  field  when vortices cease to
overlap significantly.   A minimum in $\sigma
(B)$ thus occurs due to the competition of these
broadening mechanisms  in the vicinity of the
transition to the VG phase.  %

\begin{figure}[thb]
\includegraphics[width=16cm]{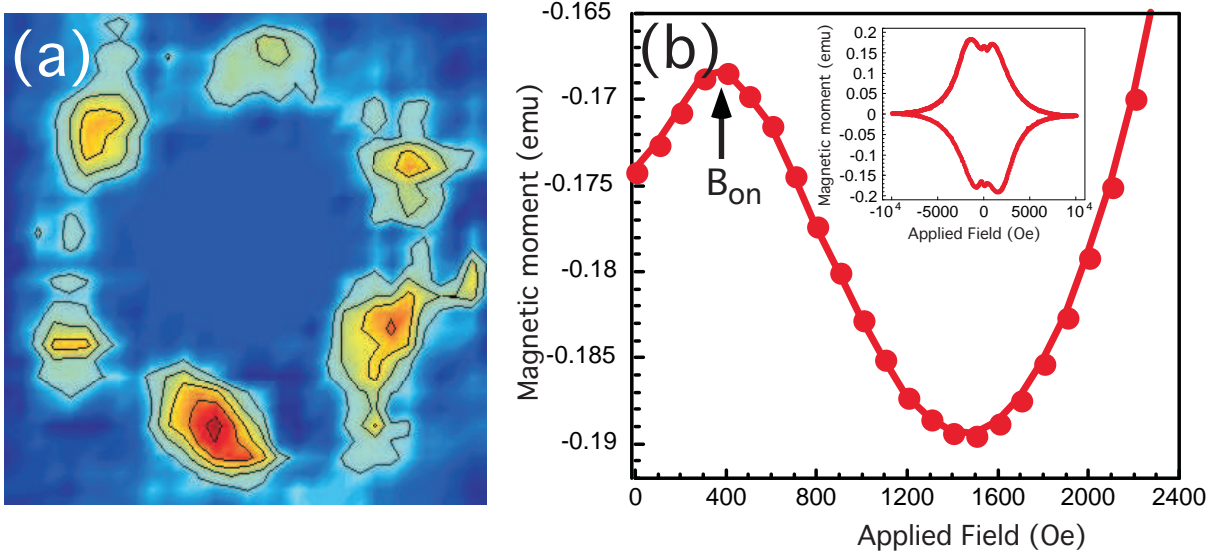}
\caption[]{\label{fig3} (Color)
a)   A small angle neutron diffraction pattern showing the  existence of an ordered vortex phase at low fields (150 Oe) and temperature (6.2 K), having hexagonal symmetry.  The intensity of this pattern rapidly falls as the field is increased towards the VG phase (see text). b) The feature in the magnetisation at 16.5~K  is  associated with the crossover to the VG state, measured on a small piece taken from the $\mu$SR sample.  The inset shows a complete hysteresis loop.}
\end{figure}

We have also recently used SANS to  perform a
preliminary experiment to image the vortex
arrangement in the same LSCO sample, using
instrument D22 at the Institut Laue-Langevin,
Grenoble, France.   At an applied field of 150~Oe
an hexagonal diffraction pattern could be
measured (Fig.~\ref{fig3}a),  thus confirming the
quasi-long-range order of the vortex lattice at
low fields.   At  a temperature of 5~K  the
intensity was found to drop very rapidly as a
function of applied field. Furthermore, the
diffraction spots  show an  appreciable
broadening in $q$  in the vicinity of $B_{cr}\sim
800$~G \cite{drew03}, which, in accord with the
$\mu$SR data, implies an increase of static
disorder in the system around this field. While
in $\mu$SR experiments a field-cooled (FC)
experiment is necessary to interpret the
lineshapes, in other systems this is known to not
always represent the most ordered state
\cite{Dewhurst}. In the SANS experiment we also
attempt to improve the lattice perfection in the
Bragg-glass region by cooling in a modulated
applied field, as in ref. \cite{Dewhurst}.  The
lattice perfection was always found to be reduced
by these approaches, indicating that the FC state
for both SANS and $\mu$SR represents the most
ordered vortex lattice state at low fields.

\begin{figure}[thb]
\includegraphics[width=16cm]{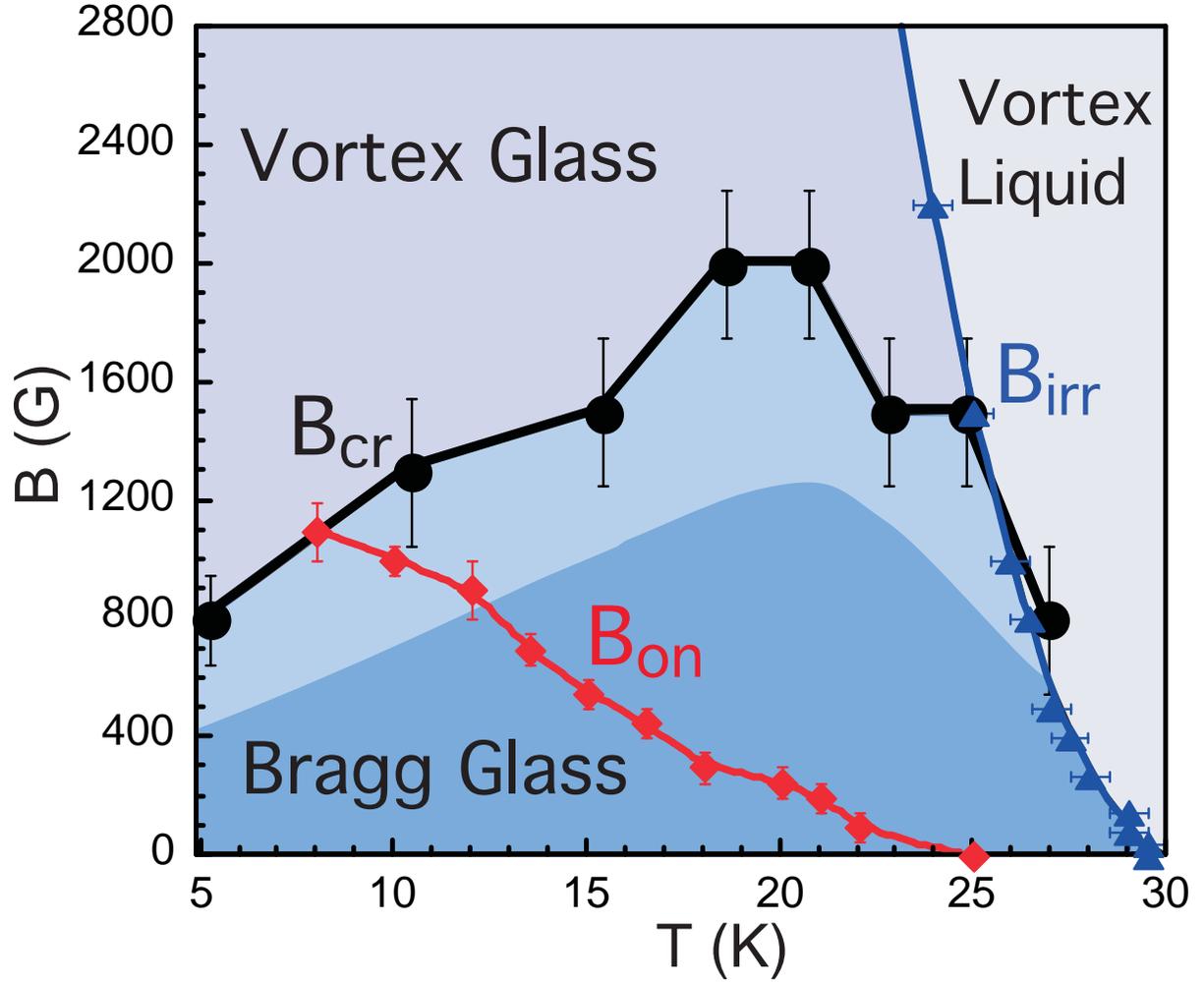}
\caption[]{\label{fig4} (Color)
The magnetic phase diagram derived from the changes observed 
in the $\mu$SR field distributions.  $B_{cr}(T)$ indicates the onset of the 
 broadening at high field, which is significant only below
$\sim 25$~K, and should be considered as an upper limit for the BG-VG transition (see text). 
This uncertainty in the exact position of the transition is represented schematically by the shading below the line $B_{cr}(T)$.  $B_{irr}(T)$ indicates the irreversibility line as determined by bulk measurements of the field-cooled/zero-field-cooled  (FC/ZFC) magnetisation using a SQUID magnetometer. The position of the feature  in the magnetisation curves $B_{on}(T)$ is also plotted (see Fig.\ref{fig3}b).  
} 
\end{figure}

In the literature, evidence for the existence of
the BG-VG transition is frequently taken from
signatures in the bulk magnetisation, since the
transition to the glassy state is accompanied by
an increased effectiveness of point pinning.  In
Fig.~\ref{fig3}b  is a section of an $M(H)$ loop
at 16.5 K,  where a `peak effect' is evident as
an anomalous increase in the magnitude of the
magnetisation beginning at around 400~Oe. The
temperature dependence of the position of the
onset of  this peak $B_{on}(T)$ is plotted in the
phase diagram of Fig.~\ref{fig4}.  Similar
features  have  previously been associated with
the BG-VG transition e.g. \cite{Giller97}. This
peak  disappears at around 25~K, as indicated  by
the line $B_{on}(T)$ of Fig.~\ref{fig4},  the
same temperature at which $B_{cr}(T)$ intersects
the irreversibility line $B_{irr}(T)$ (Fig.
\ref{fig4}).

The occurrence of an equilibrium phase transition
with increasing magnetic field  has been
discussed theoretically  by many authors
includingGiamarchi and Le Doussal
\cite{Giamarchi94}. It is worth noting in this
context that the muon is sensitive only to the
{\em local} magnetic field, so in general this
transition will be manifest as a crossover of
behaviour which reflects the underlying sharp
transition. In Fig.~\ref{fig4} the line $B_{cr}
(T)$ is a plot of   the minima in $\sigma (B) $,
where the broadening due to static glassy
disorder begins to dominate. This feature thus
provides an upper limit for the BG-VG transition.
An  upward trend of  this BG-VG transition,
reminiscent of that observed experimentally,
$B_{cr}(T)$,  has been predicted to arise from
the reduced effectiveness of pinning at higher
temperature \cite{Giamarchi94,Ertas}.  The
dependence  of the signature observed in the
magnetisation measurements, $B_{on}(T)$,  is very
differerent to $B_{cr}(T)$. This is not
surprising given that  the muons measure $p(B)$
in a field-cooled state, whereas the
magnetisation measurements determine  the
macroscopic properties of a  disturbed system
possessing strong macroscopic flux gradients, and
reflect the changing dynamic response to the
pinning landscape. The SANS intensity, by
contrast,  is sensitive both to thermal
fluctuations and to the loss of long range
coherence in the vortex structure.

In conclusion we have measured the vortex state
at a microscopic level and observed a change with
field from a BG to a VG phase, the latter
possessing only short range transverse order.
The minimum in the $\mu$SR linewidth with
increasing field, observed at $B_{cr}(T)$, arises
primarily from the competing influences  of
increasing elastic interactions combined with
thermal fluctuations, and  a rapid collapse of
translational correlation lengths in the vortex
system within the plane {\em perpendicular} to
the field direction.  This  provides unambiguous
evidence for a transition with increasing field
from a nominally Bragg glass phase to a more
disordered vortex glass state.
This is the first such measurement on a system
composed of well-coupled  vortex lines, and
paves the way for a detailed study of the
evolution of short range order in the presence of
weak pinning, a problem of universal significance
\cite{Giamarchi94}.

%
%
\acknowledgments

This work was supported by the Swiss National Science Foundation,  the Engineering and Physical Sciences Research Council of the United Kingdom, the DST (India) and the Ministry of Education, Science and Technology  of Japan. The $\mu$SR experiments were performed at the Swiss Muon Source, Paul Scherrer Institute, Villigen.

\newpage
\vspace{-3mm}
\bibliographystyle{unsrt}

\begin{thebibliography}{10}

\bibitem{Larkin79}
A. I.~Larkin, J. Low Temp. Phys. {\bf 34} 409 (1979).

\bibitem{Giamarchi94}
T.~Giamarchi and P.~Le Doussal, Phys. Rev. Lett. {\bf 72} 1530 (1994).

\bibitem{Cubitt93}
R.~Cubitt {\em et al.}, Nature {\bf 365} 407-411 (1993).

\bibitem{Mesot02}
R. Gilardi {\em et al.}, Phys. Rev. Lett. 88 (21) 217003 (2002).

\bibitem{Klein01}
T.~Klein {\em et al.}, Nature {\bf 413} 404 (2001).

\bibitem{blatterbible}
G.~Blatter {\em et al.} Rev. Mod. Phys. {\bf 66} 1125 (1995).

\bibitem{Zeldov}
E. Zeldov {\em et al.}, Nature {\bf 375} 373 (1995).

\bibitem{Schilling}
A. Schilling {\em et al.}, Nature {\bf 382} 791 (1996).

\bibitem{Pardo}
F. Pardo {\em et. al.} , Nature. {\bf 396} 348 (1998).

\bibitem{review}
C.M.~Aegerter and S.L.~Lee, Appl. Magn. Resn. {\bf 13}  75-93 (1997).

\bibitem{Blatter_em}
G.~Blatter {\em et al.}, Phys. Rev. B {\bf 54} 72 (1996).

\bibitem{Brandtlowtemp}
E.H.~Brandt, J. Low. Temp. Phys. {\bf 73} 355 (1988).

\bibitem{Menon02}
G.I. Menon, Phys. Rev. B {\bf 65} 104527 (2002).

\bibitem{Brandtpancakes}
E.H.~Brandt, Phys. Rev. Lett. {\bf 66} 3213 (1991).

\bibitem{Menon99}
G.I. Menon {\em et al.}, Phys. Rev. B {\bf 60} 7607 (1999).

\bibitem{caveat}
The onset of magnetic order can also lead to a broadening of the lineshapes.
  This effect is indeed observed for this sample, but only below 3 K, at which
  temperature both the longitudinal and transverse $\mu$SR signals indicate the
  onset of magnetic order. However, the highly broadened transverse lineshapes
  observed below 3K in the magnetically ordered regime retain the general
  characteristics of an ordered vortex line arrangement at low field, and
  indicate the transition to a disordered phase at high field. These results
  will be discussed elsewhere in more detail.

\bibitem{Brandt88}
E.H. Brandt, Phys. Rev. B {\bf 37} 2349 (1988).

\bibitem{lee93}
S.L.~Lee {\em et al.}, Phys. Rev. Lett. \textbf{71} 3862 (1993).

\bibitem{Song93}
Y.-Q.\ Song {\em et al.}, Phys. Rev. Lett. {\bf 70} (1993) 3127.

\bibitem{lee95}
S.L.~Lee {\em et al.}, Phys. Rev. Lett. \textbf{75}, 922 (1995).

\bibitem{drew03}
A.J.~Drew {\em et al.}, detailed neutron measurements to be published
  elsewhere.

\bibitem{Dewhurst}
S.J.~Levett, C.D.~Dewhurst and D.~McK.~Paul, Phys. Rev. B {\bf 66} 014515
  (2002).

\bibitem{Giller97}
D.~Giller {\em et. al.} , Phys. Rev. Lett. {\bf 79} 2542 (1997).

\bibitem{Ertas}
Denis Ertas and David R. Nelson, Physica C {\bf 272} 79 (1996).

\end{thebibliography}

\newpage

\end{document}